# *Transformation Electronics: Tailoring Electron's Effective Mass*

*Mário G. Silveirinha*[(1, 2)] *and Nader Engheta*[(1)*]

*(1) University of Pennsylvania, Department of Electrical and Systems Engineering, Philadelphia, PA, U.S.A., engheta@ee.upenn.edu*
*(2) University of Coimbra, Department of Electrical Engineering – Instituto de Telecomunicações, Portugal, mario.silveirinha@co.it.pt*

**Abstract**

The speed of integrated circuits is ultimately limited by the mobility of electrons or holes, which depend on the effective mass in a semiconductor. Here, building on an analogy with electromagnetic metamaterials and transformation optics, we describe a new transport regime in a semiconductor superlattice characterized by extreme anisotropy of the effective mass and a low intrinsic resistance to movement – with zero effective mass – along some preferred direction of electron motion. We theoretically demonstrate that such regime may permit an ultra fast, extremely strong electron response, and significantly high conductivity, which, notably may be weakly dependent on the temperature at low temperatures. These ideas may pave the way for faster electronic devices and detectors and new functional materials with a strong electrical response in the infrared regime.

---

[*] To whom correspondence should be addressed: E-mail: engheta@ee.upenn.edu

In 1969, Esaki and Tsu suggested that by either periodically doping a monocrystalline semiconductor or by varying the composition of the alloy, quantum mechanical effects should be observed in a new physical scale[1], so that the conduction and valence bands of such *superlattices* are structured in the form of many sub-bands[1,2], and in particular they predicted the possibility of a negative differential conductance.[1] This pioneering work has set the stage for the dispersion engineering in semiconductor superlattices. This conceptual breakthrough and other prior key proposals (e.g. the idea of quasi-electric fields[3]), are the foundation of many spectacular advances in semiconductor technology[4], and has enabled among others the development of the quantum cascade laser[5], and the realization of ultrahigh mobilities in semiconductor superlattices and quantum wells[6,7].

Following these advancements, more recently, there has been a huge activity in the study of a new class of mesoscopic materials – metamaterials – whose electromagnetic properties are determined mainly by the geometry and material of its constituents, rather from the chemical composition[8,9]. Such line of research has resulted in the development of double negative materials, which promise erasing diffraction effects and perfect lensing [8].

Until now, the obvious analogy between superlattices and electromagnetic metamaterials received little attention, apart from isolated studies[10,11]. Here, inspired by the exciting paradigm offered by electromagnetic metamaterials and transformation optics[8,9], we develop the paradigm of "transformation electronics", wherein the electron wave packets are constrained to move along desired paths, and predict a totally new transport regime in a semiconductor superlattice based on the extreme anisotropy of the effective mass.

In a semiconductor the effective mass determines the inertia of the electron to an external stimulus. The finite value of the mobility ultimately limits the speed of integrated circuits and other devices. In most electronic circuits the electron flow is supposed to occur along a predetermined path, e.g. down the passageway connecting two transistors. However, typically only a small portion of the available free carriers responds effectively to an external electric field, i.e. those whose velocity $\mathbf{v}_g = \hbar^{-1} \nabla_\mathbf{k} E$ is parallel to the impressed field. Would it however be possible to engineer the electron mass in such a way that all the available electronic states contribute to the electron flow? Moreover, would it however be possible to reverse or "cancel" the effects of the intrinsic electron resistance to movement, along the preferred direction of motion?

A superlattice with the properties implicit in the first question must be anisotropic. Indeed, in order that $\mathbf{v}_g = \hbar^{-1} \nabla_\mathbf{k} E$ is parallel to the desired direction of flow (let us say *z*), it is necessary that the energy dispersion $E = E(\mathbf{k})$ depends exclusively on the wave vector component $k_z$, and hence the effective mass tensor satisfies $m^*_{xx} = m^*_{yy} = \hbar^2 \left( \partial^2 E / \partial k_y^2 \right)^{-1} = \infty$, i.e. the resistance to a flow in the *x-y* plane must be extremely large. To satisfy the requirements implicit in the second question it is necessary that $m^*_{zz} = \hbar^2 \left( \partial^2 E / \partial k_z^2 \right)^{-1}$ be near zero. Thus ideally we should have $m^*_{xx} = m^*_{yy} = \infty$ and $m^*_{zz} = 0$, and thus an effective mass tensor characterized by *extreme anisotropy*. Notably, heterostructures with extreme anisotropy have received some attention in recent years due to their potentials in collimating both light[12] and electrons[13]. However, our findings are fundamentally different from previous studies: we deal with a bulk semiconductor superlattice, and show how by combining two different semiconductors it may be possible to super-collimate the electron flow ($m^*_{xx} = m^*_{yy} = \infty$) and in addition to have a weak resistance to movement ($m^*_{zz} = 0$). A zero mass has been previously predicted to occur at contacts between semiconductors with normal and inverted band structures[14], but not an extreme anisotropy regime.

To achieve this, we draw on an analogy with electromagnetic metamaterials. The intriguing tunneling phenomena observed in electromagnetic metamaterials are rooted in the fact that two materials such that $\varepsilon_1 = -\varepsilon_2$ and $\mu_1 = -\mu_2$, with $\varepsilon$ being the permittivity and $\mu$ the permeability, "electromagnetically annihilate" one another.[8,15] It is thus natural to wonder if in electronics it may be possible to identify complementary materials that when paired yield $m^* \approx 0$. Since, the effective mass of the carriers is expected to be determined by some averaging of the values of $m^*$ in the superlattice constituents, this suggests that one should look for materials wherein $m^*$ has *different signs*.

Even though unusual, the carriers can have a *negative effective mass*, notably in semiconductors and alloys with a negative energy band gap.[16] Examples of such materials are mercury-telluride (HgTe) [a group II-VI degenerate semiconductor] and some alloys of mercury-cadmium-telluride (HgCdTe), which have an inverted band structure[16,17], so that the $\Gamma_8$ (P-type) valence bands lie above the conduction band $\Gamma_6$ (S-type), and the effective masses of both electrons and holes ($m^*_{c,h}$) are negative.

In Refs. [18, 19] we develop a formal analogy between the Helmholtz equation for the electromagnetic field and a Schrödinger-type equation for the *envelope wavefunction* consistent with the standard Kane model (k·p method) for semiconductors with a zincblende structrure.[20] Within this formalism, that is consistent with Bastard's theory[22], the electron is described by a single component wavefunction, $\psi$, which may be regarded as the spatially averaged microscopic of wavefunction. This contrasts with the conventional k·p approach where the electron is described by a multi-component wavefunction.[20] For the case of Bloch waves, $\psi$ may be identified with the zero-th order Fourier harmonic of the microscopic wavefunction.[18, 19] Related averaging procedures have been considered previously in the context of electromagnetic metamaterials[21]. The wavefunction in the superlattice satisfies:

$$-\frac{\hbar^2}{2}\nabla\cdot\left(\frac{1}{m(E,\mathbf{r})}\nabla\psi\right)+\left(V(E,\mathbf{r})-E\right)\psi=0. \tag{1}$$

The *effective potential* $V(E,\mathbf{r})=E_c(\mathbf{r})=E_{\Gamma_6}(\mathbf{r})$ is determined by the energy level of the conduction band in each component of the heterostructure. Provided the effect of the spin-orbit split-off bands is negligible, the *dispersive* (energy-dependent) *mass* $m=m(E,\mathbf{r})$ of the heterostructure can be approximated by $m(E,\mathbf{r})\approx[E-E_v(\mathbf{r})]/(2v_P^2(\mathbf{r}))$, where $E_v(\mathbf{r})=E_{\Gamma_8}(\mathbf{r})$ is the valence band energy level, $m_0$ is the free-electron mass, $E_P=2P^2m_0/\hbar^2$, $P$ is Kane's parameter[20], and $v_P=\sqrt{E_P/(3m_0)}$ has dimensions of velocity. The dispersive mass, $m=m(E,\mathbf{r})$, should not be confused with the effective mass $m^*=\hbar^2\left[\partial^2 E/\partial k_i\partial k_j\right]^{-1}$ determined by the curvature of the energy diagram. For narrow gap semiconductors $m^*$ satisfies (for both electrons and holes): $m_c^*\approx m_h^*\approx\dfrac{E_g}{2v_P^2}$, where $E_g=E_c-E_v$ is the band-gap energy of the semiconductor. The sign of $m^*$ is the same as that of $E_g$.

Let us consider a superlattice formed by slabs of two narrow gap semiconductors alternately stacked along the z-direction (Fig. 1a). Each semiconductor layer ($i$=1,2) has thickness $d_i$, and is described by parameters $V_i$ and $m_i = m_i(E)$, and the band gap energies of the semiconductors have different signs so that $E_{g,1} > 0$ (e.g. an alloy of HgCdTe) and $E_{g,2} < 0$ (e.g. HgTe). In addition, the valence band offset $\Lambda = E_{v,2} - E_{v,1}$ is such that $0 < \Lambda < E_{g,1} + |E_{g,2}|$ so that there is no overlap between conduction and valence bands in the two materials (Fig. 1c). From the analogy between the Schrödinger [Eq. (1)] and Helmholtz equations outlined in[18,19], we have the correspondences between the parameters $\varepsilon$ and $\mu$ (permittivity and permeability) and $V$ and $m$: $\mu(\omega) \leftrightarrow m(E)$ and $\varepsilon(\omega) \leftrightarrow E - V(E)$, analogous to[10]. Hence, it follows that the material with positive band gap ($E_{g,1} > 0$) is seen by an electron with energy $E$ in the band gap as a material with $\varepsilon < 0$ and $\mu > 0$ (ENG material), whereas the material with negative band gap ($E_{g,2} < 0$) is seen as a material with $\varepsilon > 0$ and $\mu < 0$ (MNG material) [see Fig. 1b].

We calculated analytically the dispersion of the superlattice Bloch modes, using Eq. (1) and imposing generalized Ben Daniel-Duke boundary conditions at the interfaces[22]. Our Kronig-Penney type model yields[19],

$$\cos(k_z a) = \cos(k_{z,1} d_1)\cos(k_{z,2} d_2) - \frac{1}{2}\left(\frac{k_{z,1} m_2}{k_{z,2} m_1} + \frac{k_{z,2} m_1}{k_{z,1} m_2}\right)\sin(k_{z,1} d_1)\sin(k_{z,2} d_2) \quad (2)$$

where $a = d_1 + d_2$ is the lattice constant, $k_{z,i} = \sqrt{\frac{2m_i(E)}{\hbar^2}(E - V_i(E)) - k_\parallel^2}$, $\mathbf{k} = \mathbf{k}_\parallel + k_z \hat{\mathbf{z}}$ is the wave vector, and the effective parameters of the semiconductors are $V_i = E_{c,i}$ and $m_i = (E - E_{v,i})/2v_{P,i}^2$, $i$=1,2. The conduction mini-band resulting from the hybridization of the energy diagrams of the two semiconductors emerges at the energy level $E = V_{eff}$, where $V_{eff} = V_1 f_1 + V_2 f_2$. The energy origin is fixed so that $V_{eff} = 0$.

A detailed analysis of Eq. (2) reveals that the effective mass of the superlattice satisfies $M_{zz}^* = 0$, and $M_{xx}^* = M_{yy}^* = \infty$ (for both electrons and holes) provided the spatially averaged band-gap energy ($E_{g,av}$) and the filling ratio of the materials satisfy[19]:

$$E_{g,av} = \frac{\Lambda}{2} \frac{v_{P,1}^2 - v_{P,2}^2}{v_{P,1}^2 + v_{P,2}^2} \quad \text{and} \quad f_1 = f_2 = \frac{1}{2}. \quad (3)$$

For ternary alloys of Hg$_{1-x}$Cd$_x$Te we have $v_{P,1} \approx v_{P,2}$, because Kane's $P$ parameter varies little with the mole fraction $x$.[17], and hence Eq. (3) reduces to $E_{g,av} = 0$. This can be realized taking the negative band gap material as HgTe ($E_{g,2} = -0.3\,eV$ [16,17]), and the positive band gap material as Hg$_{0.65}$Cd$_{0.35}$Te, which has $E_{g,1} = +0.3\,[eV]$[23]. Fig. 2 shows the effective dispersive mass and effective potential calculated using our model (with $\Lambda = 0.40|E_g| = 0.12\,eV$[24]) confirming that the effective parameters of each material have different signs. In the conditions of Eq. (3), and for $v_{P,1} = v_{P,2} \equiv v_P$, the energy dispersion may be approximated by[19]:

$$E = \pm \hbar |k_z| v_P \left(1 + k_\parallel^2 4\hbar^2 v_P^2 / \Lambda^2\right)^{-1/2}. \quad (4)$$

Thus, the energy dispersion along the $z$-direction varies linearly, consistent with the property $M_{zz}^* = 0$. Hence, even though our system is fully three-dimensional and the wavefunction is not a pseudo-spinor as in graphene, the electron transport along $z$ may be somewhat analogous to that in graphene. On the other hand, close to the surface $k_z = 0$, $\nabla_k E$ is parallel to $\hat{z}$, and thus all the associated electronic states contribute to an electron flow along the z-direction, as expected from $M_\parallel^* = \infty$. These properties are confirmed by Fig. 3a [obtained using Eq. (2)] which depicts the energy dispersion for an Hg$_{0.65}$Cd$_{0.35}$Te- HgTe superlattice with $a = 6a_s = 3.9$nm, with $a_s = 0.65$nm the lattice constant of the bulk semiconductors.[17] Fig. 3c shows that the dispersion calculated with Eq. (4) captures accurately the results of the Kronig-Penney model. The value of $v_P$ in the superlattice is similar to that of the Fermi velocity in graphene, $v_P = \sqrt{E_P / 3m_0} = 1.06 \times 10^6 m/s$ [$E_p = 19 eV$ [17]], and hence, in the limit of low scattering the electron response in the superlattice can be extremely fast. Similar to photonic metamaterials a description of the superlattice in terms of effective parameters is possible when $ka \ll 1$, where $k$ represents the wave vector in a generic region. The spread of the wave vector is determined by the temperature, and thus at low temperatures the effective medium theory is expected to be quite accurate. Based on Eq. (4), imposing $E \sim k_B T$ and $ka < 0.1\pi$, one may estimate that the lattice constant should not be greater than $a_{max} \sim \hbar v_P 0.1\pi / (k_B T)$, which at room temperature gives $a_{max} \sim 8nm$.

Ideally, the energy dispersion should be independent of $k_\parallel$. This is achieved close to the plane $k_z = 0$, where the constant energy surfaces are flat, whereas for larger values of $k_z$ they become hyperbolic (Fig. 3b). Indeed, within the validity of Eq. (4), the ideal case requires $\Lambda \to \infty$. In Fig. 3d, it is shown that if the lattice constants of the materials are slightly mismatched, the energy dispersion is perturbed and a small band gap may appear.

Even in this non-ideal scenario, the effective mass $M_{zz}^*$ remains near zero, whereas $M_{\parallel}^*$ remains extremely large (not shown).

The transport properties of the superlattice, and most notably the conductivity, may be radically different from those of the constituent semiconductors. A detailed calculation shows that within the validity of Eq. (4), the intraband conductivity is given by[19]

$$\sigma_{\text{intra,xx}} = \sigma_{\text{intra,yy}} = \frac{ie^2}{\hbar^2 \omega} \frac{1}{\hbar v_P} \frac{1}{6} (k_B T)^2 \left( D + \frac{1}{D} - 2 \right) \qquad (5a)$$

$$\sigma_{\text{intra,zz}} = \frac{ie^2}{\hbar^2 \omega} \frac{1}{\hbar v_P} \left( \frac{\Lambda}{2\pi} \right)^2 (D - 1) \qquad (5b)$$

where $D = \sqrt{(k_{\parallel,\max} 2\hbar v_P / \Lambda)^2 + 1}$ and $k_{\parallel,\max} \sim \pi/a_s$ is a cut-off parameter.[19] It is assumed that the Fermi level lies exactly at $E = 0$. Moreover, we neglect scattering effects due to defects or interface mismatch, which may in any case be modeled phenomenologically by replacing $\omega$ by $\omega + i\Gamma$ in the above formulas, where $\Gamma$ represents a collision frequency. It should be mentioned the effective medium model based on Eq. (1) is unable to predict the dispersion of (the hybridized) heavy-hole states, and thus their contribution to the conductivity was not considered. However, since the heavy-hole mini-bands are expected to be nearly flat they should not influence much the transport properties.

Equation (5) predicts that the conductivity along the $z$-direction is independent of the temperature in the regime $M_{zz}^* = 0$ and $M_{\parallel}^* = \infty$. Furthermore, the conductivity may be characterized by extreme anisotropy, and when $D \gg 1$ the anisotropy ratio $\sigma_{\text{intra,zz}} / \sigma_{\text{intra,xx}} = 3/(2\pi^2)(\Lambda/k_B T)^2$ may be extremely large. Even though Eq. (5) was derived using the approximate dispersion (4), Fig. 4a shows that it describes fairly well the conductivity calculated using Eq. (2) (the case $\Lambda = 0.4 E_g$ models the superlattice

Hg$_{0.65}$Cd$_{0.35}$Te-HgTe). Due to the extreme anisotropy, at low temperatures most of the states contribute to the electron flow along z, and thus the conductivity in the *x-y* plane vanishes in the limit $T \to 0$. The anisotropy ratio, as well as the absolute value of $\sigma_{zz}$, are enhanced with larger values of $\Lambda$, because larger values of $\Lambda$ yield an energy dispersion closer to the ideal case: $E = \pm \hbar |k_z| \text{v}_P$. Fig. 4b and 4c show that the superlattice conductivity can be made several orders of magnitude ($\sim 10^3$ at $T = 300K$) larger than that of the constituent materials. To shed light on the intriguing independence of $\sigma_{zz}$ on *T*, using Eq. (4) we calculated the density of states in the superlattice, $g(E) = \frac{1}{6\pi^2} \left( \frac{\Lambda}{2} \frac{1}{\hbar \text{v}_P} \right)^2 \frac{1}{\hbar \text{v}_P} (D^3 - 1)$, which is independent of the energy and does not vanish at the Fermi level (Fig. 4d). Hence, at the Fermi level the surfaces of constant energy are not reduced to a point as in a normal semiconductor (or graphene), but instead are collapsed into the $k_z = 0$ plane (a square-shaped surface). The electrons occupying such states can respond effectively to an external field oriented along *z*, which explains the finite conductivity in the $T \to 0$ limit.

In conclusion, we have investigated the transport properties of a novel metamaterial-inspired superlattice, characterized by linear energy dispersion along some preferred direction of carrier motion and extreme anisotropy. The condition $M^*_{zz} = 0$ results from pairing materials with band-gaps of different signs that effectively interact as "matter-antimatter", in the same manner as ENG and MNG metamaterials electromagnetically annihilate one another. Our ideas may establish a new paradigm for an ultra-fast and extremely strong electronic response, which may be nearly independent of temperature in the limit $T \to 0$, and exciting new developments in electronics and photonics. As the concept of "Transformation Optics" enables tailoring the path of light, in our system we can have the same but for electrons, namely, the electron's path may be constrained so that the electrons are forced to move along a preferred direction. Since it may be possible to vary the parameters of semiconductors continuously either by doping or by controlling the material composition or – in case of a 2D electron gas – by tailoring externally the potential seen by the electrons with a top-gate, we envision that some of the ideas of Transformation Optics can be brought to the field of electronics.

*Figures*

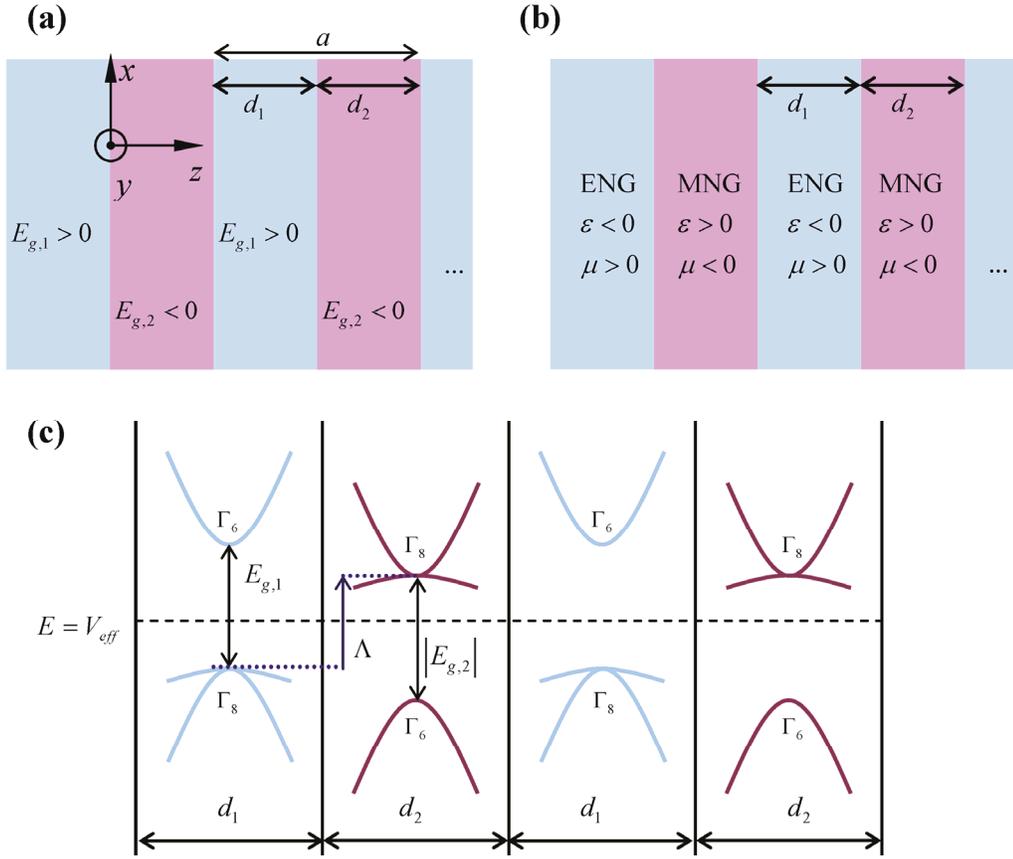

Fig. 1. **Transformation Electronics and Electronic Metamaterials**: Sketch of the geometry and electronic band diagram of the elements of the superlattice. (a) Geometry of a stratified superlattice formed by alternating layers of semiconductor alloys with band gaps with different signs. (b) Electromagnetic analogue of the superlattice for energy levels close to $E - V_{eff} \approx 0$: in the band gap the semiconductor with positive (negative) band gap is the electronic analogue of a $\varepsilon < 0$ ($\mu < 0$) electromagnetic material. (c) Detailed energy band structure of each layer of the superlattice, showing the valence band-offset $\Lambda$ between the two semiconductors.

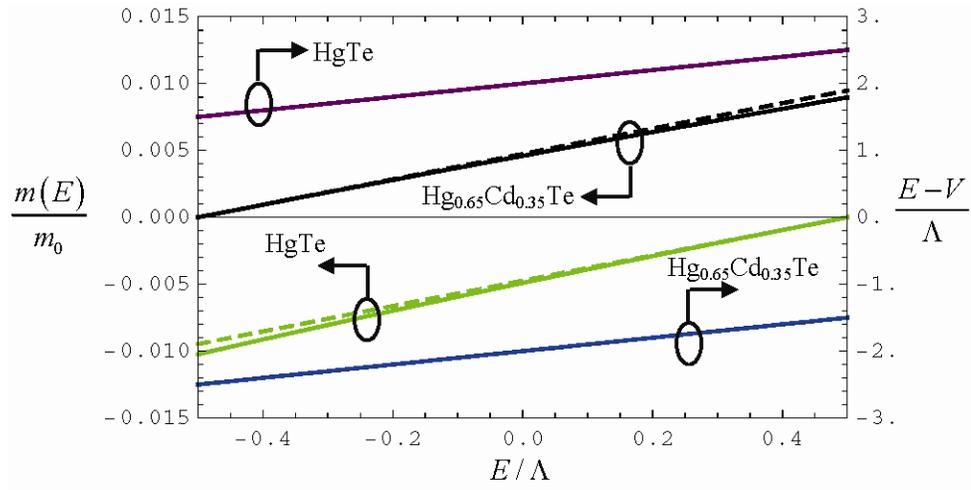

Fig. 2. **Effective parameters (mass and potential) of the semiconductor alloys**. Left axis: dispersive mass as a function of the normalized electron energy $E$; Solid lines: exact result taking into account the effect of the split-off bands (see Ref. [18, 19]); Dashed lines: linear mass approximation described in the main text. As seen, the effect of the split-off bands in negligible in the energy range of interest. Right axis: effective potential ($E-V$) as a function of the normalized electron energy $E$.

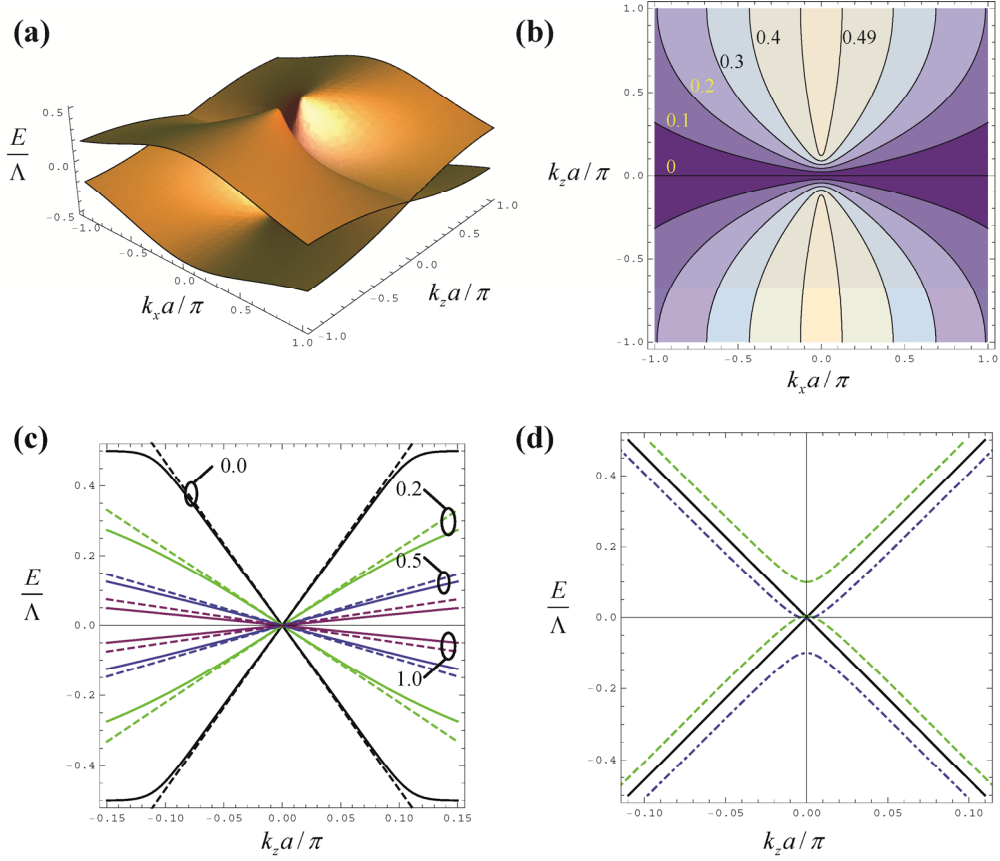

Fig. 3. **Electronic band structure of the superlattice electronic metamaterials**. (a) Energy dispersion of a superlattice with $v_P = 1.06 \times 10^6 m/s$, $E_{g1} = -E_{g2} = 0.30 eV$, $\Lambda = 0.40 E_{g1}$, $f_1 = f_2 = 0.5$ and $a = 6a_s = 3.9 nm$. (b) Contours of constant energy in the *x-z* plane. (c) Comparison between the energy dispersion calculated using the Kronig-Penney model [Eq. (2)] and the approximate result given by Eq. (4) for the values of $k_x a/\pi = 0.0, 0.2, 0.5, 1.0$. (d) Energy dispersion of the superlattice when: $f_1 = f_2 = 0.5$ (black solid line); $f_1 = 0.48$ (green dashed line); $f_1 = 0.52$ (blue dot-dashed line);

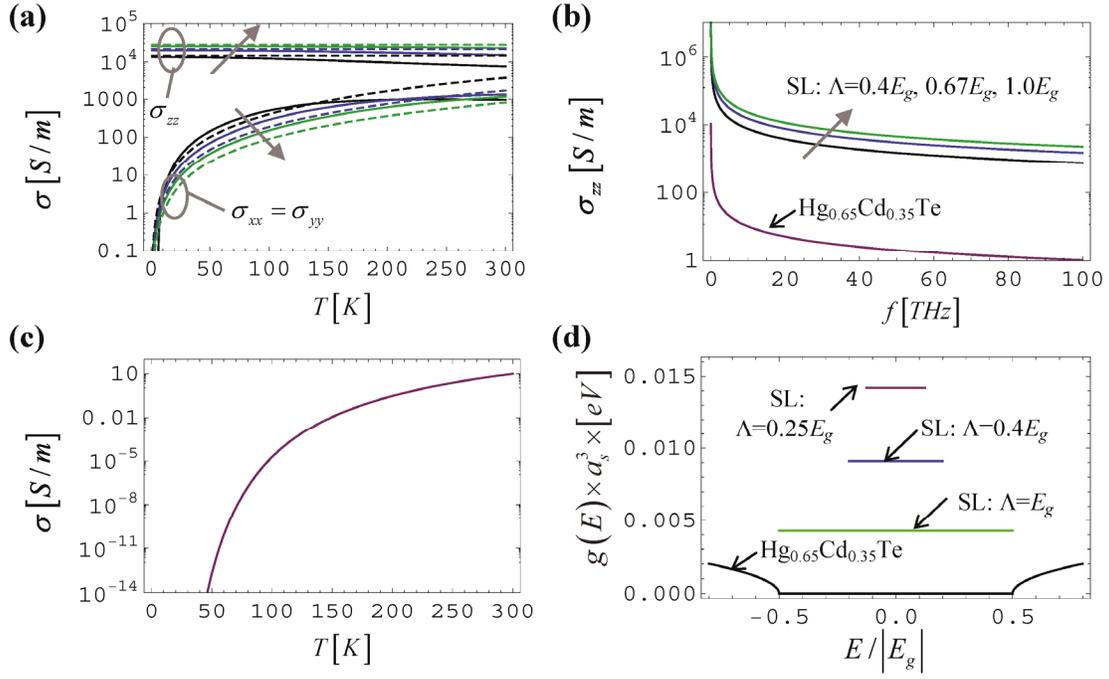

Fig. 4. **Conductivity and density of states of the superlattice.** (a) Conductivity of the semiconductor superlattice (SL) at 10THz as a function of the temperature for different values of the valence band offset: $\Lambda = E_g$ (green lines), $\Lambda = 0.67 E_g$ (blue lines), and $\Lambda = 0.4 E_g$ (black lines). The gray arrows indicate the direction of increasing $\Lambda$. The solid lines were calculated using the "exact" energy dispersion of the superlattice, whereas the dashed lines were obtained from Eqs. (5a)-(5b). (b) Conductivity of the semiconductor superlattice as a function of frequency for different values of the valence band offset at 300K. (c) Similar to (a) but for the normalized intraband conductivity of $Hg_{0.65}Cd_{0.35}Te$, assuming that the Fermi level lies at the midpoint of the energy band gap. (d). Normalized density of states of the semiconductor superlattice for different values of the valence band offset. In all the calculations it was assumed that $k_{\parallel,\max} = 0.1\pi / a_s$. Since $g(E)$ was computed based on the approximate model (4), the results of Fig. 4d are meaningful only for $|E| \ll \Lambda$.